\documentclass[11pt]{article}
\usepackage[dvips]{graphicx}  
\begin{document}

\title{\vspace{20mm}
\LARGE \bf Synthetic Quantum  Systems}  
\author{{Reginald T. Cahill\footnote{{\bf Process Physics URL:}
scieng.flinders.edu.au/cpes/people/cahill\_r/processphysics.html}}\\
  {School of Chemistry, Physics and Earth Sciences}\\{ Flinders University
}\\ { GPO Box
2100, Adelaide 5001, Australia }\\{\small(Reg.Cahill@flinders.edu.au)}\\  \\
Published in {\it Smart Materials and Structures}  {\bf 11}, 699-707(2002). }

\date{}
\maketitle

\begin{abstract}
So far proposed quantum computers use fragile and environmentally sensitive natural quantum
systems. Here we explore the new notion that synthetic quantum systems suitable for quantum
computation may be fabricated from smart nanostructures using topological excitations of a
stochastic neural-type network that can mimic natural quantum systems. These developments are a
technological application of process physics which is an information theory of reality
in which space and quantum phenomena are emergent, and so indicates the deep origins of quantum
phenomena.  Analogous complex stochastic dynamical systems have recently been proposed within
neurobiology to deal with the emergent complexity of biosystems, particularly the biodynamics of
higher brain function.  The reasons for analogous discoveries in fundamental physics and
neurobiology are discussed.
\end{abstract}




\newpage
\section{Introduction}
\label{sect:intro} 

Fundamental discoveries  in science have always resulted in accompanying technological
developments, not only through the possibilities  provided by  new phenomena, materials and 
processes but, perhaps most important of all, through the change in mindset driven by     these
discoveries. Here I explore new technological possibilities related to the development of, essentially, a
quantum theory of gravity; a new theory that unifies the phenomena of space and time with quantum
phenomena. This synthesis has arisen not through yet another  extension of our current mindset
but by the development of a profoundly different way of comprehending and modelling reality at its deepest
levels.  The follow-on new technology relates to the possibility of synthetic quantum systems and their
use in a new class of quantum computers. The characterisation of this class suggests, at this very early
stage, that these will be unlike both conventional classical and currently envisaged quantum computers,
but will have many characteristics in common with biological neural networks, and may be best suited for
artificial intelligence applications.  Indeed the realisation that the phenomena of synthetic quantum
systems is possible may amount to a discovery also  relevant to our understanding of biological neural
networks themselves.   But first we must review the nature of and the need for fundamental changes of the
mindset prevailing in the physical sciences. These changes relate to the discovery that we
need to comprehend reality as a complex semantic information system which only in part can be
approximately modelled by syntactical information systems.   

Physics, until recently,  has always used syntactical information systems to model
reality. These are a development of the Euclideanisation of geometry long ago.  Such systems begin with
undefined `objects', represented by symbols, together with {\it a priori} rules of manipulation of these
symbols; these rules being a combination of mathematical manipulations together with `the laws of physics'
expressed in mathematical notation.  This is the game of logic. But logic has only a limited connection to
the phenomena of time, for logic is essentially non-process: it is merely symbol manipulation. In physics
time has always been modelled by geometry. This non-process model matches the notion of order, but fails
to match the notion of past, present and future. For this reason physics has always invoked a metarule to
better match this  model with the experienced phenomena of time. This metarule involves us imagining a
point moving at uniform rate along the geometrical time line.  Despite the success of this model, its
limitations have led to enormous confusion in physics and elsewhere, particularly when reality and models
of reality are not clearly distinguished.  For example it is often asserted that time {\it is}
geometrical; it {\it is} the 4th dimension.  The more successful a model is the more likely is this
confusion to arise; and also the stronger is the urge to resist an examination of failures of this model.

One consequence of the refusal to examine other models, particularly of time, 
has been the failure to find a model that unifies the phenomena of space, time and
the quantum. As well limitations of self-referential syntactical information 
systems were discovered by G\"{o}del\cite{G}.  The famous incompleteness theorem 
asserts that there exists truths which are unprovable within such an information
system.   Chaitin\cite{Chaitin} has demonstrated that the unprovable truths
arise from a structural randomness in that there is insufficient structure for
such truths to be compressed into the axioms of the system. This structural randomness
is related to but is outside  a non-process syntactical system. This is different from the
randomness observed in quantum measurement processes, which is randomness of events in
time; nevertheless there are suggestive similarities. The quantum
randomness is also beyond the syntactical formalism of the quantum theory;
quantum theory is described by entirely deterministic
mathematics  (which is by its nature is non-process) and the randomness is invoked via the Born  metarule.
As for the geometrical model of time, these metarules are outside of the syntax, but must not
be inconsistent with it. 

The analogies between the quantum measurement randomness and the structural randomness 
associated with self-referential syntax systems has suggested that reality may be a self-referential
system subject to intrinsic informational limitations.  This has led to the development of the 
{\it process physics} modelling of reality
\cite{CK97,CK98,CK99,CKK00,C01,C02,CK,RC03,RC04}, see also \cite{MC}. This model uses a non-geometric
process model for time, but   also  argues for the importance of 
relational or semantic information in modelling reality. Semantic information refers to
the notion that reality is a purely informational system where the information is
internally meaningful.  The study of a pure semantic information system is   by means of 
a subtle bootstrap process. The mathematical model for this has the form of a stochastic
neural network. Non-stochastic neural networks  are well known for their
pattern recognition abilities\cite{neural}.  The stochastic behaviour is related to the
limitations of syntactical systems, but also results in the neural network being
innovative in that it creates its own patterns.  The neural network  is
self-referential, and   the stochastic input,  known as self-referential noise (SRN),
acts both to limit the depth of the self-referencing   and also to generate
potential order.
Such information has the form of
self-organising patterns which also generate their own `rules of interaction'.   Hence the
information is `content addressable', rather than is the case in the usual syntactical
information modelling where the information is represented by symbols.  

  In the process physics space and quantum physics are
emergent  and unified, and time is a distinct non-geometric process.  Quantum
phenomena are caused by fractal topological   defects  embedded in and
forming a growing three-dimensional fractal process-space. This amounts to the discovery of
a quantum gravity model.  As discussed in Refs.\cite{C01, C02}  the emergent physics  includes
limited causality, quantum field theory, the Born quantum measurement metarule, inertia,
time-dilation effects, gravity and the equivalence principle, and black holes, leading in
part to an induced Einstein spacetime phenomenology. In particular this new physics predicts
that Michelson dielectric-mode interferometers can reveal absolute motion relative to the quantum
foam which is space. Analysis \cite{CK,RC03,RC04} of existing experimental dielectric-mode
interferometer  data confirms that absolute motion is meaningful and measurable.  These are
examples of an effective syntactical system being induced by a  semantic system.

Here we explore one technological development which essentially follows as an application 
of the  quantum theory of gravity.  The key discovery has been  that self-referentially
limited  neural networks display quantum behaviour.  It had already been noted by
Peru\v{s}\cite{Perus} that the classical theory of non-stochastic neural networks\cite{neural}
had similarities with deterministic quantum theory, and so this development is perhaps not unexpected,
atleast outside physics.  This suggests that artificial or synthetic quantum systems (SQS)
may be produced by a stochastic self-referentially limited  neural network, and later we
explore how this might be achieved technologically. This possibility could lead to the
development of robust quantum computers.   However an even more intriguing insight becomes
apparent, namely that the work that inspired the classical theory of  biological neural
networks   may have overlooked the possibility that sufficiently complex
biological  neural networks may in fact be essentially quantum computers. This possibility has
been considered by Kak\cite{Kak} and others, see works in Pribram\cite{Pribram}. 

As well we draw attention to the work of Freeman {\it et al} \cite{Freeman} where it is argued
that biocomplexity requires the development of new mathematical models having the form of complex
stochastic dynamical systems driven and stabilised by noise of internal origin through
self-organising dynamics. These conclusions are based on extensive work on the biodynamics of
higher brain function. Analgous equations here, see Eqn.(1), also arise but by using entirely
different arguments  dealing with and addressing deep problems in the traditional modelling
of reality within physics. Of course the common theme and emerging understanding is that both
reality and biocomplexity entail the concept of internal or semantic information, and presumably
there are generic aspects to this, which it now seems, have been independently discovered within
physics and neurobiology.  Of course dynamical systems like (1) dealing with reality as a whole
presumably entail and subsume the phenomena of biocomplexity.

\section{Self-Referentially Limited  Neural Networks}
\label{sect:NN}

\begin{figure}
\setlength{\unitlength}{0.5mm}
\hspace{20mm}
\begin{picture}(150,60)
\thicklines
\put(10,10){\circle{15}}\put(8,7){1}
\put(30,30){\circle{15}}\put(27,27){2}
\put(48,0){\circle{15}}\put(45,-3){3}
\put(15,15){\vector(1,1){10}}\put(30,12){$B_{23}>0$}
\put(5,15){\vector(-1,1){10}}
\put(12,3){\vector(1,-4){3}}
\put(35,25){\vector(1,-2){9.5}}
\put(33,37){\vector(1,2){7}}
\put(37,32){\vector(3,1){15}}
\put(55,3){\vector(3,1){15}}
\put(34,-15){\vector(1,1){9}}
\put(95,15){\circle{15}}\put(94,12){i}
\put(100,20){\vector(1,1){10}}
\put(90,20){\vector(-1,1){10}}
\put(25,-22){(a)}   \put(115,-22){(b)}    \put(210,-22){(c)}

\put(145,15){\circle{15}}\put(144,12){i}
\put(150,20){\vector(1,1){10}}
\put(140,20){\vector(-1,1){10}}
\put(95,9){\oval(8,20)[b]}
\put(96,-1){\vector(1,0){1.0}} 
\put(115,15){$\in$}
\end{picture}
\hspace{40mm}
\setlength{\unitlength}{0.20mm}

\hspace{105mm}
\begin{picture}(0,50)(40,0)  
\thicklines
\put(155,165){\line(3,-5){60}}
\put(155,165){\line(-3,-5){60}}
\put(115,100){\line(3,-5){42}}
\put(195,100){\line(-3,-5){21}}
\put(135,160){ \bf $i$}
\put(225,160){ \bf $D_0\equiv 1$}
\put(225,100){ \bf $D_1=2$}
\put(225,60){ \bf $D_2=4$}
\put(225,25){ \bf $D_3=1$}
\put(155,165){\circle*{5}}
\put(115,100){\circle*{5}}
\put(195,100){\circle*{5}}
\put(95,65){\circle*{5}}
\put(135,65){\circle*{5}}
\put(175,65){\circle*{5}}
\put(215,65){\circle*{5}}
\put(155,30){\circle*{5}}
\end{picture}

\caption{\small (a) Graphical depiction of the neural network with links
$B_{ij}\in {\cal R}$ between nodes or pseudo-objects. Arrows indicate sign of
$B_{ij}$. (b) Self-links are internal to a node, so $B_{ii}=0$. (c) An $N=8$ spanning
tree with $L=3$.  The  distance distribution $D_k$ is indicated for node {\it i}.
\label{section:figure:neural}}
\end{figure}
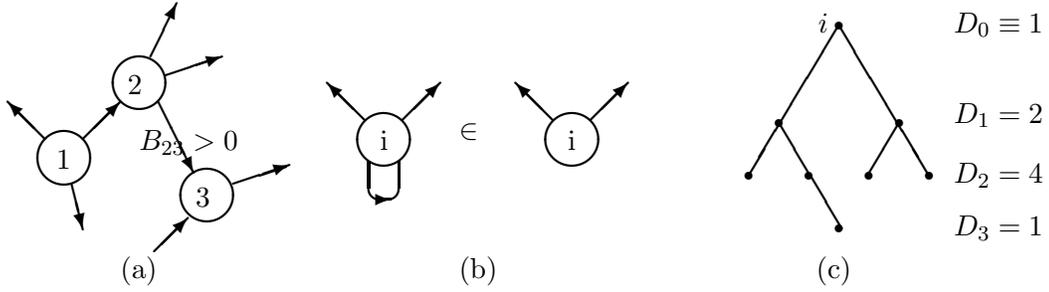

Here we briefly describe a model for a  self-referentially limited neural network and in the
following section we describe how such a network results in emergent quantum behaviour, and which,
increasingly, appears to be a unification of space and quantum phenomena. Process physics is  a semantic
information system and is devoid of {\it a priori} objects and their laws  and so it requires a subtle
bootstrap mechanism to set it up. We use a stochastic neural network, Fig.1a, having the structure of 
  real-number valued connections or relational information strengths $B_{ij}$ (considered as forming a
square matrix) between  pairs of nodes or pseudo-objects
$i$ and $j$ . In standard 
neural networks \cite{neural} the network  information resides in both link and node
variables,  with the semantic information residing in attractors of the iterative network.
Such systems are also not pure in that there is an assumed underlying and manifest {\it a
priori} structure.
  
 The nodes and their link variables  will be revealed  to be themselves sub-networks of informational
relations. To avoid explicit self-connections
$B_{ii}\neq 0$, which are a part of the sub-network content of
$i$, we use antisymmetry $B_{ij}=-B_{ji}$ to conveniently ensure that 
$B_{ii}=0$, see Fig.1b.

At this stage we are using a syntactical system with symbols $B_{ij}$ and, later, rules for the
changes in the values of these variables. This system is the syntactical seed for the pure
semantic system.   Then to ensure that the nodes and links are not remnant {\it a priori}
objects the system must generate strongly linked  nodes (in the sense that the $B_{ij}$ for
these nodes are much larger than the $B_{ij}$ values for non- or weakly-linked nodes) forming
a fractal network; then self-consistently the start-up nodes and links may themselves be
considered   as mere names for sub-networks of relations.  For a successful suppression  the scheme must
display self-organised criticality (SOC)\cite{SOC} which acts as a filter for the start-up syntax. The
designation `pure' 
 refers to the notion that all seeding syntax has been removed. SOC is the process
where the emergent behaviour  displays universal criticality in that the  behaviour is
independent of the individual start-up syntax; such a start-up syntax  then  has no 
ontological significance.  

To generate a fractal structure we must  use a non-linear iterative system for the
$B_{ij}$ values.  These iterations amount to the  necessity to introduce a time-like
 process.   Any system possessing {\it a priori}  `objects' can never
be fundamental as the explanation of such objects must be outside  the system.  Hence in
process physics the absence of intrinsic undefined objects is linked with the phenomena of
time, involving as it does an ordering of `states', the present moment effect, and the
distinction between past and present. Conversely in non-process physics the presence of {\it a
priori } objects is related to the use of the non-process geometrical model of time, with this modelling
and its geometrical-time metarule being an approximate emergent description from process-time.  In this
way process physics arrives at a new modelling of time, {\it process time}, which is much more complex
than that introduced by Galileo, developed by Newton, and reaching its high point with Einstein's
spacetime geometrical model. 

The  stochastic neural network   so far has been realised with one
particular  scheme involving a stochastic non-linear matrix iteration.  The matrix
inversion $B^{-1}$ then models self-referencing in that it requires  all elements of $B$ to
compute any one element of $B^{-1}$. As well there is the  additive SRN  
$w_{ij}$ which limits the self-referential information  but, significantly, also acts in such
a way that the network is innovative in the sense of generating semantic information, that is
information which is internally meaningful.  The emergent behaviour is believed to be
completely generic in that it is not suggested that reality is a computation, rather it
appears that reality is essentially very minimal and having the form of an order-disorder
information system.  

To be a successful contender for the Theory of Everything (TOE) process
physics must ultimately prove the uniqueness conjecture:  that the  characteristics (but not
the contingent details) of the  pure  semantic information system are unique.  This would
involve demonstrating both the effectiveness of the SOC filter and the robustness of the
emergent phenomenology, and the complete agreement of the later with observation.    

The stochastic neural network is  modelled by the iterative process
\begin{equation}
B_{ij} \rightarrow B_{ij} -\alpha (B + B^{-1})_{ij} + w_{ij},  \mbox{\ \ } i,j=1,2,...,2M;
M
\rightarrow
\infty.
\label{eq:map}\end{equation}
 where 
 $w_{ij}=-w_{ji}$ are
independent random variables for each $ij$ pair and for each iteration, chosen from some probability
distribution. Here $\alpha$ is a parameter the precise value of which should not be critical but which
influences the self-organisational process. 
We start the iterator at 
$B\approx 0$, representing the absence of information.  
  With the noise absent the iterator would converge in a deterministic and reversible manner to
 a constant matrix.
However in the presence of the noise the iterator process  is non-reversible and non-deterministic. 
It   is also manifestly non-geometric and non-quantum, and so does not assume any of the
standard features of syntax based physics models. 
The dominant  mode is  the  formation of a randomised and structureless  background (in $B_{ij}$).
However   this noisy iterator also manifests  a   self-organising process which results in a growing 
three-dimensional  fractal process-space that competes with this random background - the formation of a
`bootstrapped universe'.  The emergence of order in this system might appear to violate
expectations regarding the 2nd Law of Thermodynamics; however because of the SRN the system
behaves as an open system and the  growth of order arises from selecting implicit order in the
SRN.  Hence the SRN acts as a source or negentropy, and the need for this can be traced back
to G\"{o}del's incompleteness theorem.

 This  growing  three-dimensional  fractal process-space is an example of a Prigogine far-from-equilibrium
dissipative structure driven by the SRN. 
 From each iteration the noise term
will additively introduce rare large value
$w_{ij}$.  These  $w_{ij}$, which define sets of linked nodes, will persist   through more
iterations than smaller valued $w_{ij}$ and, as well,  they become further linked  by the iterator
to form a three-dimensional process-space with embedded topological defects.
 
 To see this consider a node $i$ involved in one such large $w_{ij}$;   
 it will be   connected via  other large $w_{ik}$ to a
number of other nodes and so on, and this whole set of connected nodes forms a connected random graph unit
which we call a gebit as it acts as a small piece of geometry formed from random information links and 
from which the process-space is self-assembled. The gebits compete for new links  and undergo
mutations. Indeed, as will become clear, process physics is remarkably analogous in its operation to
biological systems. The reason for this is becoming clear: both reality and subsystems of reality must use
semantic information processing to maintain existence, and symbol manipulating systems are totally unsuited
to this need, and in fact totally contrived.

To analyse the connectivity of such 
gebits assume for simplicity that the large $w_{ij}$  arise with fixed but very small probability $p$,
then the geometry of  the gebits is revealed by studying the probability distribution for  the structure
of the random graph units or gebits minimal spanning trees with $D_k$ nodes  at $k$ links from node $i$
($D_0
\equiv 1$), see Fig.1c, this is given by\cite{Nagels}

\begin{equation}{\cal P}[D,L,N] \propto \frac{p^{D_1}}{D_1!D_2!....D_L!}\prod_{i=1}^{L-1}
(q^{\sum_{j=0}^{i-1}{D_j}})^{D_{i+1}}(1-q^{D_i})^{D_{i+1}},\end{equation}
where $q=1-p$, $N$ is the total number of nodes in the gebit and $L$ is the maximum depth from node $i$. 
To find the most likely connection pattern we numerically maximise ${\cal P}[D,L,N]$ for fixed $N$ with respect
to
$L$ and the $D_k$. The resulting $L$ and $\{D_1,D_2,...,D_L\}$ fit very closely to the form $D_k\propto
\sin^{d-1}(\pi k/L)$;  see Fig.2a  for $N=5000$ and $\mbox{Log}_{10}p=-6$.  The resultant  $d$
values for a range of $\mbox{Log}_{10}p$ and $N=5000$ are shown in Fig.2b. 

\vspace{-35mm}
\hspace{-20mm}\includegraphics{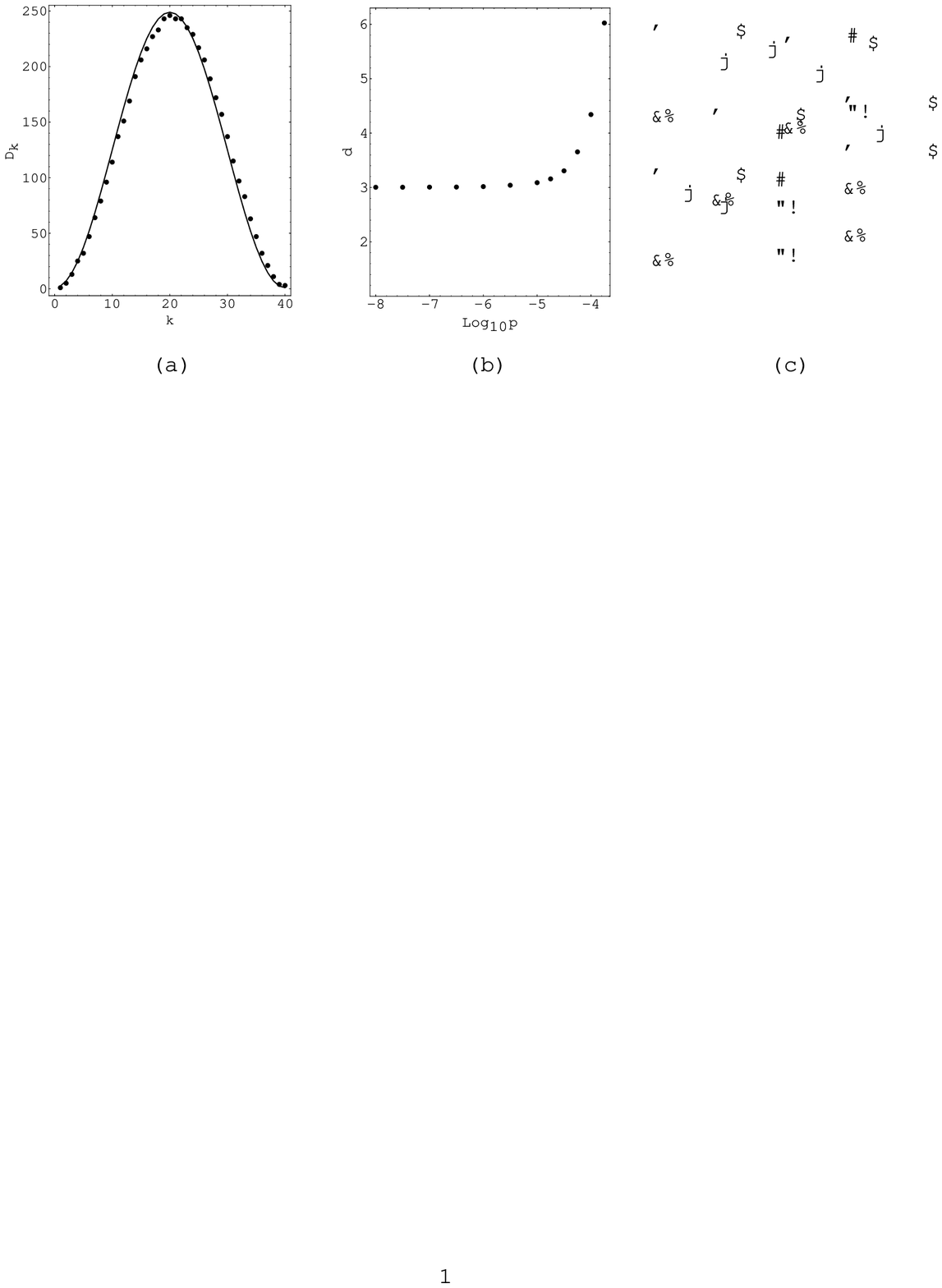}

\vspace{70mm}
\begin{figure}[h]
\caption{\small 
(a) Points show the $D_k$ set and $L=40$ value found by numerically  maximising ${\cal P}[D,L,N]$
for $\mbox{Log}_{10}p=-6$ for fixed  $N=5000$. Curve shows
$D_k\propto \sin^{d-1}(\frac{\pi k}{L})$ with best fit $d=3.16$ and $L=40$, showing  excellent
agreement, and indicating  embeddability in an $S^3$ with some topological defects. (b) Dimensionality $d$ of
the gebits as a function of  the probability $p$. (c) Graphical depiction of the `process space' at one stage of the iterative
process-time showing a quantum-foam structure formed from  embeddings and linkings of gebits.  The
linkage connections have the distribution of a 3D space, but the individual gebit components are
closed compact spaces and cannot be embedded in a 3D background space.  So the drawing is only
suggestive. Nevertheless this figure indicates that process physics generates a cellular
information system, where the behaviour is determined at all levels by internal information. } 
\end{figure}

This shows, for $p$ below a critical value, that
$d=3$ indicating that the connected nodes have a natural embedding in a 3D hypersphere $S^3$; call this
a base gebit. Above that value of $p$,   the increasing value of $d$
indicates the presence of extra links that, while some conform with the embeddability,  are in the main defects
with respect to the geometry of the $S^3$.  These extra links act as topological defects.  By themselves these
extra links will have the   connectivity and embedding geometry of numbers of gebits, but these gebits have a
`fuzzy' embedding in the base gebit. This is an indication of  fuzzy  homotopies  (a homotopy is,
put simply, an embedding of one space into another).  

The base gebits $g_1, g_2, ...$ arising from the SRN together with their embedded topological defects have
another remarkable property:  they are `sticky' with respect to the iterator.  Consider the   larger valued 
$B_{ij}$ within a given gebit  $g$, they form  tree graphs and most tree-graph adjacency matrices are singular
 (det$(g_{tree})=0$).  However  the presence of other smaller valued $B_{ij}$ and the general
 background noise  ensures that det$(g)$ is small  but not exactly zero. 
Then the 
$B$ matrix has an inverse with large components that act to cross-link  the new and
existing gebits.  If this cross-linking was entirely random then the above analysis  could again be used and we
would conclude that  the base gebits themselves are formed into a 3D hypersphere with embedded topological
defects.  The nature of the resulting 3D process-space is suggestively indicated in Fig.2c. 

Over ongoing
iterations the existing gebits become cross-linked and eventually lose their ability to undergo further
linking; they lose their `stickiness'  and decay.  Hence the emergent space is 3D but is continually
undergoing  replacement of its component gebits;  it is an informational process-space, in sharp distinction to
the non-process continuum geometrical spaces that have played a dominant role in modelling physical space.  If
the noise is `turned off' then this emergent dissipative space will decay and cease to exist.  We thus see that
the nature of space is deeply related to the logic of the limitations of logic, as implemented here as a
self-referentially limited neural network.

\section{Emergent Quantum Behaviour}

Relative to the iterator the dominant  resource is the large valued $w_{ij}$ from the SRN
because they form the `sticky' gebits which are self-assembled into the non-flat compact 3D process-space. The 
accompanying topological defects within these gebits and also the topological defects within the process space
require a more subtle description.  The key behavioural mode for those defects which are  sufficiently large
(with respect to the number of component gebits) is that their existence, as identified by their topological
properties, will survive the ongoing process of mutation, decay and regeneration; they are topologically
self-replicating.   Consider the analogy of a closed loop of string containing a knot - if, as the string ages,
we replace small sections of the string by new pieces then eventually all of the string will be replaced;
however the relational information represented by the knot will remain unaffected as  only the topology of the
knot is preserved.
  In the process-space there will be gebits embedded in gebits, and so forth,  in topologically
non-trivial ways;  the topology of these embeddings is all that will be self-replicated in the processing of
the dissipative structure.  
\label{sect:SQS} 

To analyse and model the life of these topological defects we need to characterise
their general behaviour: if sufficiently large (i) they will self-replicate if topologically non-trivial,
(ii)  we may apply continuum homotopy theory to tell us which embeddings are topologically
non-trivial,  (iii)  defects will only dissipate if embeddings of `opposite winding number' (these classify the
topology of the embedding)  engage one another, (iv)  the embeddings will be in general
fractal, and (iv) the embeddings need not be `classical', ie the embeddings will  be fuzzy.  To track the 
coarse-grained behaviour of such a system has lead  to the development of  a new form of quantum field
theory:  Quantum Homotopic Field Theory (QHFT).  This models both the
process-space and the topological defects.
QHFT has the form of an iterative functional Schr\"{o}dinger equation for the discrete time-evolution
of a wave-functional 
$\Psi(....,\pi_{\alpha\beta},....,t)$
\begin{equation}\Psi(....,\pi_{\alpha\beta},....,t+\Delta
t)= \Psi(....,\pi_{\alpha\beta},....,t)
-iH\Psi(....,\pi_{\alpha\beta},....,t)\Delta t +  \mbox{QSD terms},
\end{equation}      where the configuration space is that of all possible 
   homotopic mappings; $\pi_{\alpha\beta}$ maps from $S_\beta$
to  $S_\alpha$ with $S_\gamma \in  \{S_1,S_2,S_3,...\}$ the set of all possible gebits (the topological defects
need not be $S^3$'s). The time step $\Delta t$ in Eqn.(3)  is relative to the scale of the fractal
processes being explicitly described, as we are using a configuration space of prescribed gebits.  At
smaller scales we would need a smaller value of   $\Delta t$.  Clearly this invokes a (finite)
renormalisation scheme. Eqn.(3), without the QSD term,  would be called a `third quantised' system in
conventional terminology. Depending on the `peaks' of 
$\Psi$ and the connectivity of the resultant dominant mappings such mappings are to be interpreted as
either embeddings  or links; Fig.2c then suggests the dominant process-space form within
$\Psi$ showing both links and embeddings. The emergent process-space then has the characteristics of a  
quantum foam. Note that, as indicated in Fig.2c, the original start-up links and nodes are now absent.
Contrary to the suggestion in Fig.2c, this process space cannot be embedded  in a {\it finite}
dimensional geometric space with the emergent metric preserved, as it is composed of infinitely nested
or fractal finite-dimensional closed spaces.  The form of the Hamiltonian $H$ can be derived from
noting that the emergent network of larger valued $B_{ij}$ behaves analogous to a non-linear elastic
system, and that such systems  have a skymionic description\cite{Manton2}; see Ref.\cite{C01}
for discussion.

There are  additional Quantum State Diffusion\cite{QSD} (QSD) terms which are non-linear and
stochastic; these QSD terms are ultimately responsible for the emergence of
classicality via an  objectification process, but in particular  they produce
wave-function(al) collapses during quantum measurements. 
The iterative functional Schr\"{o}dinger system can be given a more familiar functional
integral representation for
$\Psi$, if we ignore the QSD terms. Keeping the QSD leads to a functional integral representation for a
density matrix formalism, and this amounts to a derivation of  the decoherence process which is usually
arrived at by invoking the Born measurement metarule. Here we see that `decoherence'  arises from the
limitations on self-referencing.
In the above we have  a deterministic and unitary
evolution, tracking and preserving topologically encoded information, together with the stochastic
QSD terms, whose form protects that information during localisation events, and which
also  ensures the full matching in QHFT of process-time to real time: an ordering of
events, an intrinsic direction or `arrow' of time and a modelling of the contingent
present moment effect.   So we see that process physics generates a complete theory of quantum
measurements involving the  non-local, non-linear and  stochastic QSD terms.  It does this because
it generates both the `objectification' process associated with the classical apparatus and the actual
process of (partial)  wavefunctional collapse as the quantum modes interact with the measuring
apparatus.  Indeed many of the mysteries of quantum measurement are resolved when it is realised that
it is the measuring apparatus itself that actively provokes the collapse, and it does so because the
QSD process is most active when the system deviates strongly from its dominant mode, namely the
ongoing relaxation of the system to a 3D process-space.  This is essentially the process that
Penrose\cite{Penrose} suggested; namely that the quantum measurement process is essentially a
manifestation of quantum gravity. The demonstration of the validity of the Penrose argument of course
could only come about when  quantum gravity was {\it derived} from deeper considerations, and not by
some {\it ad hoc} argument such as the {\it quantisation} of Einstein's classical spacetime model.  Again
we see that there is a direct link between G\"{o}del's theorem on the limitations of self-referencing
syntactical systems and the quantum measurement process.

The mappings  $\pi_{\alpha\beta}$ are related to group manifold parameter spaces with the group determined by
the dynamical stability of the mappings. This gauge symmetry leads to  the flavour symmetry of the
standard model. Quantum homotopic mappings or skyrmions  behave as fermionic or bosonic modes for 
appropriate winding numbers; so process physics predicts both fermionic and bosonic quantum modes, but
with these  associated with topologically encoded information and not with  objects or `particles'.

\section{Quantum Computers and Synthetic Quantum Systems}
\label{sect:QC}
In previous sections was a description of  a fundamental theory of reality that uses a
self-referentially limited neural network scenario inspired by the need to implement subsymbolic
semantic information processing, resulting, in particular, in a quantum theory of gravity. The neural
network  manifests complex connectionist patterns that behave as  linking and embedded
gebits; with some forming topological defects. The latter are essentially the quantum modes that
at a higher level have been studied as quantum field theory. Rather than  interfering with this
emergent quantum mode behaviour the intrinsic noise (the SRN) is a {\it sine qua non} for its
emergence; the SRN is a source of negentropy that refreshes the quantum system. Here I discuss the
possible application of this effect to the construction of quantum computers (QC) which use Synthetic
Quantum Systems (SQS). 

Quantum Computers\cite{Nielsen}  provide the means for an exponential speed-up effect compared to
classical computers, and so offer technological advantages for certain, albeit so far restricted,
problems.  This speed-up results from the uniquely quantum characteristics of entanglement and the
quantum measurement process, both of which now acquire explanation from process physics.   This
entanglement allows the parallel unitary time evolution, determined by a time-dependent programmed 
hamiltonian,  of superpositions of states, but results in the `output' being encoded in the complexity of
the final wavefunction; the encoding is that of the amplitudes of this wavefunction when expanded into
some basis set.  These individual amplitudes are not all accessible, but via  a judicious choice of
quantum measurement, determined by  the problem being studied,  the required information is accessed.

In these early days  the construction of simple quantum computers all use naturally occurring quantum
systems, such as interacting individual atoms embedded in a silicon lattice.  These are both
difficult to construct and sensitive to environmental noise. They will play a key role in 
testing the concepts of quantum computation.  The decoherence caused by this environmental noise can be
partly compensated by error-correction codes\cite{Nielsen}.  However as first noted by
Kitaev\cite{Kitaev,Freedman} quantum computers that use topological quantum field systems would achieve a
fault-tolerance by virtue of the topological excitations being protected from local
errors.  From process physics we now see that this process is also the key to the very
origin of quantum behaviour. 

This all suggests that a robust and large scale quantum computer might ultimately be best constructed
using a stochastic neural network architecture which exhibits topological quantum field behaviour; a
synthetic quantum system (SQS).  Such a SQS would essentially manifest synthetic entanglement and
synthetic collapse to achieve the apparent advantages of quantum computation, and the inherent
robustness would follow from the non-local character of topological modes.  Indeed the actual
`information' processing would be achieved by the interaction of the topological modes.  There are
 a number of key questions such as (i) is such a SQS  indeed possible?
(ii)  how could such a system could be `constructed'? (iii) how  could it be programmed? (iv) how
is information  to be inserted and extracted?  I shall not tackle here the fundamental question
of whether the phenomenon of a SQS is possible, for to answer that question will require a long and 
difficult analysis.  But assuming the final answer is yes, we shall try here to at least characterise such
a system. 

 We first note that such a synthetic quantum system based quantum computer may be very
different in its area of application compared to the quantum computers being currently considered. This
new class of QC would best be considered as being non-symbolic and non-algorithmic; indeed it might best
be characterised as a semantic information computer. The  property of being non-symbolic follows from
the discussions in the previous sections in relation to reality itself, and that information in a SQS
QC has an internal or semantic meaning.  Semantic information or  knowledge, as it might best be
described, would be stored in such a QC in the form of non-local topological states which are
preserved by the topological character of such states, although more static and primitive
information could be preserved in the `classical' structure of the neural network.  Such a SQS would
be driven by essentially the negentropy effect of thermal noise, with the effective `iterator' of
the system selecting special patterns from  this noise; this noise essentially acting as a pseudo-SRN.  So
a SQS QC is essentially a `room temperature'  QC which, combined with its topological features, would make
such a system inherently robust.  The noise also manifests a synthetic wave-functional collapse
mechanism, essentially a synthetic-QSD term in the time evolution equation analogous to Eqn.(3).
The effect of these collapses is that above a certain threshold the superposition property, namely
the Hamiltonian term in Eqn.(3), would be over-ridden by a non-local and non-linear collapse process; it
is from this process that the SQS QC would be inherently non-algorithmic, since the collapse is inherently
random in character. Of course as in the more `conventional' QC this non-algorithmic `measurement'
process may be exploited to extract useful and desired outcomes of the `computation'.  This
directed outcome can only be achieved if the exact character of the `measurement' can be
programmed, otherwise  the collapse will result in novel and creative outcomes; essentially the
generalisation   and linking of semantic information already in the QC. For this reason this
class of QC may represent a form of artificial intelligence, and may be best suited to 
generalisation and creativity, rather than just achieving the exponential speed-up for certain
analytical problems such as number factoring. 

It is unlikely that such a synthetic quantum system computer would be fabricated by conventional
`directed' construction technologies either old or new.  First such a computer would
necessarily have an enormous number of components in order to achieve sufficient complexity and
connectedness that topological quantum modes could be emergent.  There also needs to be
plasticity so that the collapse process can result in permanent changes to the neural network
connections, for it must be remembered that the SQS  does not operate by means of symbol
manipulation, but by the interaction and self-interaction of internal states that arise from
actual connectivity patterns within the network. It is also not clear in what manner the SQS would be
manifested. Whereas for reality itself the previous sections suggested that  connection
was sufficient, in the SQS we could also envisage the possibility that the connectivity is
manifested by other modes such as pulse timing or signal phasing.  For these reasons such a
SQS QC would probably be constructed by means of the self-assembly of enormous numbers of active
components, with node and linkage elements, whose individual survival depends of their involvement or
activity level. That is, the system could be overconnected initially and then SQS status achieved by a
thinning out process. To have such an enormous number of components then stipulates that we need very
small components and this suggests nanoscale chemical or molecular electronic self-assembly procedures.
Indeed the close connection with biological neural networks suggests that we are looking  at a
nanobiology approach, and that the self-organisation process will be biomimetic.

The programming of such a SQS QC also would have to be achieved by subtle and indirect means;
it is very unlikely that such a system could be programmed by the preparation of the initial
connection patterns or indeed by an attempt to predetermine the effective Hamiltonian for any
specified task.  Rather, like the construction of the SQS, the programming would be achieved
by means of plasticity and the biasing of internal interactions; that is via essentially biased
self-programming.   This is obvious from the role of semantic information in such a SQS; these
systems decide how information is represented  and manipulated, as the memory process is
essentially content addressing. That is, the information is accessed by  describing aspects of the
required information until  the appropriate topological patterns are sufficiently excited and
entangled that a collapse process is activated.  To bias such self-programming means that the SQS
must have essentially complex sensors that import and excite generic pattern excitations, rather
than attempt to describe the actual connectivity.  Indeed the very operation of such a SQS appears
to involve an level of ignorance of its internal operation.  If we attempt to directly probe its
operation we either observe confused and unintelligible signalling or we cause collapse  events that 
are unrelated to the semantic information embedded in the system.  Rather the best we can probably
do is exchange information by providing further input and monitoring the consequent output, and
proceed in an iterative manner.

\section{Non-Symbolic Neural Networks and Consciousness}

That a self-referentially limited neural network approach is perhaps capable of providing a deep
modelling of reality should not come as a surprise, nor is it in itself mystical or perplexing.  With
hindsight we can now say how ``else could it have been?''  Physics for some time has been moving in the
direction that reality is somehow related to information, and various `informational interpretations' of
quantum theory were advanced; but the information here, as encoded in the wavefunction,  was always
thought to be about the observer's knowledge of the state of the system, and in itself may have had
no ontological significance.  For this reason the early discoveries of quantum theory were quickly
interpreted as amounting to limits on the observer's knowledge of `particles' such as where they are
and what momentum they have, but rarely whether such point-like entities actually existed.  In this way
physicists hung to some of the oldest western science ideas of matter being `objects' in a `geometrical
space'.  Quantum phenomena were telling us a different message, but one which has been ignored for some
75 years.  Of course reality cannot be about objects and their laws; that  methodology is only suitable
for higher level phenomenological descriptions, and its success at these levels has misled us about its
suitability at deeper levels. 

 The nature of
reality  must be internally meaningful and always self-processing, the stuff of reality continues
to exists not because the `production process' is finished, but because  these
systems are self-perpetuated by this self-processing;  at this deepest level there are no
prescribed laws or entities.  The closest analogy to this idea of `internally meaningful'
information is that of the  semantic information of the human mind as experienced through our
consciousness. The processing of this semantic information is massively parallel and by all
accounts non-local.  Consciousness involves as well memory, self-modelling and self-referencing
and a shifting focus of attention.  The recognition of consciousness as a major scientific problem
has finally  occurred, and the subject is attracting  intense debate and speculation.  It is
suggested here that the difficulty science has had in dealing with this phenomena is that western
science has been entrenched in a {\it non-process} modelling of reality, and trapped in the
inherent limitations of syntax and logic. However process physics is indicating that reality is 
non-syntax and  experiential;  all aspects of reality including space itself are `occasions of
actual experience', to use  Whitehead's phrasing\cite{Whitehead,ProcessPhilosophy}.  Only information
and processes that are internally meaningful can play any role in reality; reality is not about symbols and
their syntax, although they have pragmatic use for observers.   

So process physics reveals reality to be, what is called, panexperiential.  To the extent that
these  processes result in characteristic and describable outcomes we have emergence of non-process 
syntactical language.  But in general the processes correspond to our experiences of time, and
match in particular the experience of the `present moment' or the `now'; although because of
the subtleties of communication in such a system it is not known yet whether the historical
records of separated individuals can be uniquely correlated or labelled. This is the problem
of establishing simultaneity that Einstein first drew to our attention.   In process physics
it has not yet been determined whether or not the non-local processes, say those associated
with EPR connections, enable the determination or not of an absolute frame of reference and
so absolute simultaneity.

The panexperientialism  in process physics suggests that at some level of complexity of
emergent systems that such  systems may be self-aware, not of the individual
sub-processes, but of generalised processes at the higher level; because the processes
at this level amount to self-modelling and to other characteristics of consciousness.
Such experiences cannot occur in a symbol manipulating system.                    

In process physics we see that one of the key emergent and characterising modes in a semantic
information system    is  quantum behaviour.  It is suggested that such a behaviour may also arise
in a sufficiently complex synthetic neural network subject to pseudo-SRN, resulting in synthetic quantum
systems.  Of course one system that may be manifesting such behaviour is that of our  
brains.  Simple neural networks evolved to deal with the processing and identification of various
external signals, particularly those essential to the survival of the system.  These advantages would
result in species of systems with ever larger neural networks, all behaving in the  classical mode. 
But with increasing complexity a new phenomena may have emerged, namely that of the synthetic quantum
system, particularly if its `semantic information processing' was considerably enhanced, as now
 quantum computation theory is suggesting.  Hence we are led to the speculative suggestion
that `mind' may be emergent synthetic quantum system behaviour.   That mind my
be connected to quantum behaviour has been considered by many, see Refs.\cite{Kak,Pribram,Jibu}. 
In particular the enormous number of synapses  in the dendritic network\cite{Mel} has
attracted much speculation.  Of particular relevance is that the synapses are noisy components. From
the viewpoint of synthetic quantum systems this synaptic noise would behave as a pseudo-SRN and so a
source of negentropy.

\section{Conclusions}

We have explored here some novel technological spin-offs that might conceivably arise from the
development of the quantum theory of gravity in the new process physics.   This process physics views
reality as a self-referentially limited neural network system that entails semantic information growth
and processing, and it  provides an explanation for much  if not ultimately all of the fundamental
problems in physics.  In particular process physics provides an explanation for the necessity of quantum
phenomena, and suggests that such phenomena may emerge synthetically whenever the conditions are
appropriate. These conditions may include those of the noisy neural networks that form in part our
brains, and the emergent synthetic quantum system behaviour may in fact be what we term our `mind'. But
at the technological level emergent synthetic quantum system behaviour may ultimately lead to the
development of semantic information or knowledge processing 	quantum computers which exploit the enhanced
processing possible in synthetic quantum systems with synthetic entanglement and a synthetic quantum
measurement process.  Indeed, because of the similarity of these quantum computers with our biological
neural networks and their possible manifestation of synthetic quantum behaviour, this new class of
quantum computers may display strong artificial intelligence if not even consciousness.   These
synthetic quantum system  computers may need to be essentially `grown' rather than constructed by
laying down fixed structures. To achieve this we would expect to  mimic biological neural networks,
since by arising naturally they probably represent the simplest manifestation of the required
effect.  Such a strong AI quantum computer would thus represent a major technological target for the
emerging field of nanotechnology, and indeed it would constitute  a  `smart'
 nanostruture.  While reality
demands a fractal structure for the deep reasons discussed in the text, synthetic quantum computers need
not be fractal, atleast as regards their manifest structure and operation.  Of course being a part of
reality they share in the deep underlying fractal system.

\end{document}